\documentclass[11pt]{article}

\PassOptionsToPackage{table,dvipsnames}{xcolor}
\usepackage[final]{acl}

\usepackage{times}
\usepackage{latexsym}
\usepackage{iftex}

\usepackage[T1]{fontenc}

\usepackage[utf8]{inputenc}

\usepackage{microtype}

\usepackage{inconsolata}

\usepackage{xcolor}
\usepackage[most]{tcolorbox}

\usepackage{tikz}
\usepackage{nicematrix}

\usepackage[utf8]{inputenc}
\usepackage{booktabs,multirow}
\usepackage{enumitem}
\usepackage{kotex}
\ifPDFTeX\else
\fi
\usepackage{color}
\usepackage{rotating}
\usepackage{amsmath}
\usepackage{amsfonts}
\usepackage{pifont}
\usepackage{fontawesome5}
\usepackage{array}
\usepackage{amssymb}
\usepackage{algpseudocode}
\usepackage{algorithm}
\usepackage{float}
\usepackage{placeins}
\usepackage{multicol}
\usepackage{lipsum}
\usepackage[normalem]{ulem}
\usepackage{soul}
\usepackage{listings}
\usepackage{amsmath}
\usepackage{adjustbox}
\usepackage{graphicx} 
\usepackage{colortbl} 
\usepackage{mdframed}
\usepackage{hyperref}
\usepackage{url}
\usepackage{xurl}

\tcbuselibrary{listings}

\colorlet{goalAcolor}{cyan!18}
\colorlet{goalBcolor}{magenta!14}
\newcommand{\goalhl}[2]{{\setlength{\fboxsep}{1pt}\colorbox{#1}{#2}}}

\newcommand{\ie}{{\it i.e.}}
\newcommand{\eg}{{\it e.g.}}

\newcommand{\argmax}{\operatornamewithlimits{argmax}}

\newcommand{\ours}{SteerGR}

\title{SteerGR: Steering Generative Recommendation via Sparse Autoencoders for Controllable Game Discovery}

\author{Sunkyung Lee$^1$\thanks{Work done during internship at Roblox}, Xiaokai Wei$^2$, Da Long$^2$, Hyunsik Jeon$^2$, Jiajun Wu$^2$, Rithwik Seth$^2$, \\
\textbf{Jongwuk Lee}$^1$, \textbf{Michelle Gong}$^2$, \textbf{Gungor Polatkan}$^2$ \\
$^1$Sungkyunkwan University, Republic of Korea, $^2$Roblox Corporation, USA \\
$^1$\texttt{\{sk1027, jongwuklee\}@skku.edu}, $^2$\texttt{\{xwei, dlong, hjeon, jiajunwu, rseth,} \\
 \texttt{mgong, gpolatkan\}@roblox.com}
}

\begin{document}
\raggedbottom
\clubpenalty=10000
\widowpenalty=10000
\maketitle
\begin{abstract}
Generative recommendation aims to predict item identifiers from user interaction histories, but explicit control over the properties of recommended items remains underexplored.
A key challenge is to connect human-readable concepts to internal features whose interventions effectively change the recommended items.
To address this, we propose \textit{\textbf{Steer}able \textbf{G}enerative \textbf{R}ecommender (\textbf{\ours{}})}, a framework that uses sparse autoencoders for concept-level control in game discovery.
\ours{} proceeds through three stages: (i) \textit{Concept Definition and Grounding} for defining play concepts with an LLM; (ii) \textit{Feature Discovery and Selection} for discovering candidate features by their concept associations and selecting one by its steering effect; and (iii) \textit{Sparse Feature Intervention} for amplifying or suppressing selected features to steer recommendations without retraining the recommender.
Extensive experiments with two industrial generative recommendation models on a real-world game dataset show that the selected features successfully steer recommendations toward or away from target concepts.
We believe this work takes a first step to extend SAE-based steering from LLMs to industrial generative recommendation.

\end{abstract}

\section{Introduction}
\label{sec:introduction}

Generative recommendation (GR) has emerged as a new paradigm that predicts item identifiers from a user's interaction history~\cite{nips23tiger, icml24HSTU}.
By formulating recommendation as conditional generation, GR enables the use of generative architectures and training techniques for personalized recommendation~\cite{Geng0FGZ22P5}.
Recently, GR has gained attention in industrial applications beyond academic research~\cite{sigir26snapGR, corr26netflixGR}.
As they become part of practical pipelines, understanding and controlling their behavior becomes increasingly important.

\begin{figure}[t]
\centering
\includegraphics[width=1.0\linewidth]{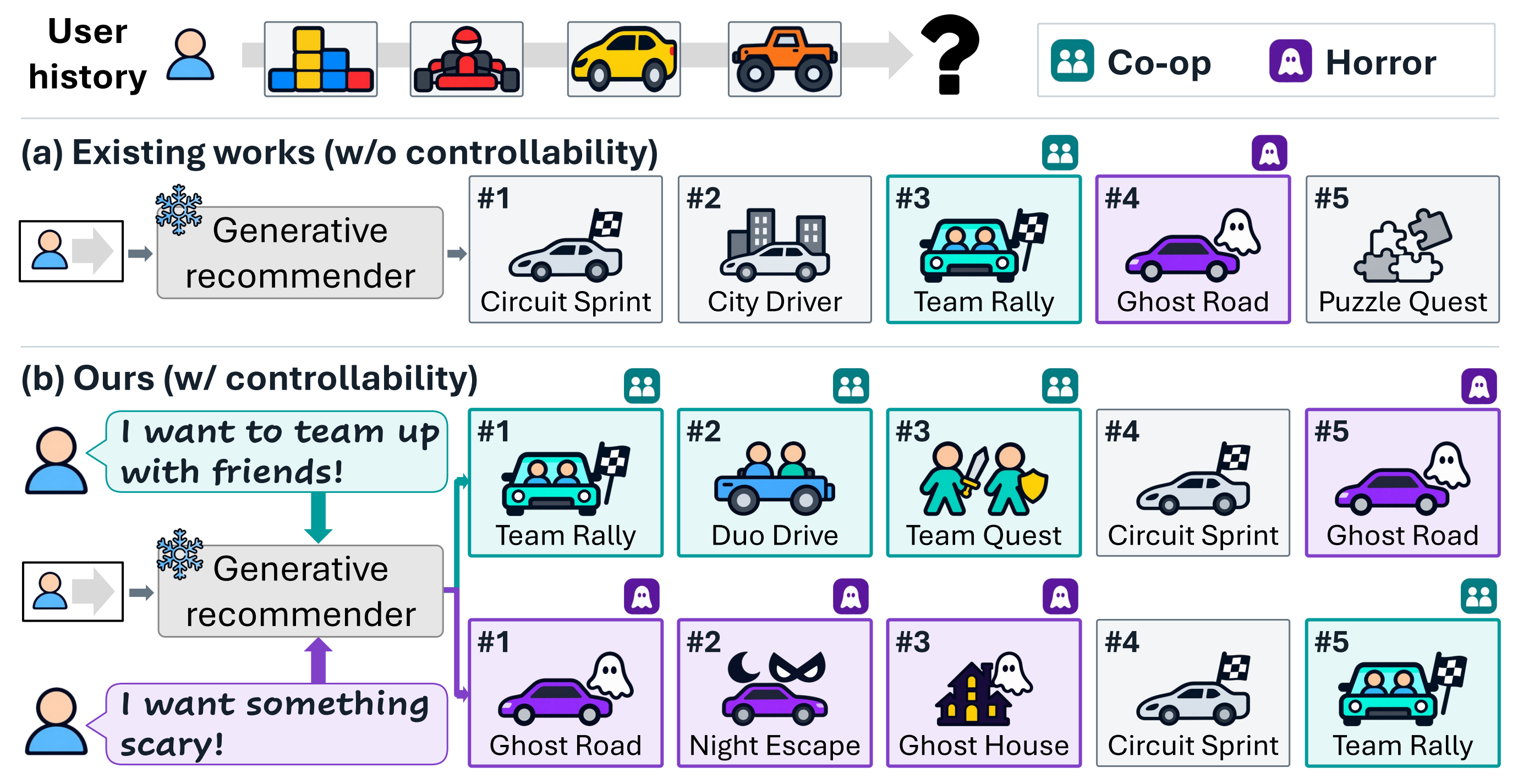}
\caption{Motivation for controllable recommendation. (a) Without explicit control, a recommender may return the fixed results regardless of the user's goals. (b) Ours can steer recommendations toward discovery goals, \eg, teaming up with friends or trying a scary experience.}\label{fig:motivation}
\vspace{-4.5mm}
\end{figure}

Beyond predicting likely interactions, recommendation services should support control toward both \textit{objectives of platforms} and \textit{discovery goals of users}~\cite{recsys15controlrs, ecweb16controlsurvey, tors25controlsurvey}. Platform operators may need to reduce exposure to particular content types, while users may seek experiences not fully reflected in their interaction histories.
As illustrated in Figure~\ref{fig:motivation}, a recommendation model without explicit control may continue to recommend racing games based on a user's interaction history (Figure~\ref{fig:motivation}a).
Controllability would instead let the user specify goals, \eg, \textit{`team up with friends'} or \textit{`try a scary experience'}, and adjust how strongly they influence recommendations conditioned on the interaction history (Figure~\ref{fig:motivation}b).
This calls for a way to connect text-defined discovery goals to internal controls in a recommendation model.

Recently, sparse autoencoders (SAEs) have been actively used to interpret internal representations of LLMs and steer generated outputs~\cite{iclr24sae,iclr25saeOpenAI,corr26saeclaude3sonnet}.
SAEs decompose dense hidden states into sparsely activated features, providing units whose responses can be examined and modified~\cite{icml25SAEBench, emnlp25SAEsurvey}.
Related studies have also applied SAEs to recommendation representations and the control of item attributes~\cite{AAAI26Monosemanticity4Rec, tois26RecSAE, www26PopSteer}.
However, connecting text-defined discovery goals to sparse feature interventions on GR remains underexplored.

To this end, we propose \textit{\textbf{Steer}able \textbf{G}enerative \textbf{R}ecommender (\textbf{\ours{}})}, a framework for concept-guided steering in generative recommendation.
The core idea is to ground text-defined play concepts in game descriptions and select the sparse features that shift recommendations toward each concept.
\ours{} proceeds through three stages.
First, \textit{Concept Definition and Grounding} scores games against LLM-defined play concepts using their descriptions, then aggregates the scores over interaction histories.
Second, \textit{Feature Discovery and Selection} contrasts SAE responses between users with high and low concept scores to identify candidates, then selects the feature that most increases the share of games with the concept on the validation split. 
Third, \textit{Sparse Feature Intervention} amplifies or suppresses the selected feature at inference time without updating the recommender parameters.
Using industrial recommenders trained on real-world game datasets, we extensively evaluate how each intervention shifts recommendations and how it affects accuracy.

Our contributions are summarized as follows:
\begin{itemize}[leftmargin=*,itemsep=2pt,topsep=3pt]
    \item \textbf{First SAE steering for GR:}
    To the best of our knowledge, we propose the first framework that connects text-defined play concepts to SAE features for steering generative recommenders.
    
    \item \textbf{Concept-guided feature discovery:} We identify an effective steering feature for each concept by contrasting user histories based on concept scores and selecting by steering effect.
    
    \item \textbf{Industrial validation:} We validate \ours{} on real-world game datasets, examining steering effects on deployed generative recommenders.

\end{itemize}

\section{Related Work}

\subsection{Generative Recommendation (GR)}
GR aims to directly generate item IDs conditioned on user interaction histories~\cite{nips23tiger}.
Early studies mainly explored effective item identifiers~\cite{Geng0FGZ22P5} and the alignment of language and collaborative semantics~\cite{ZhengHLCZCW24LCRec,wang2024letter}.
Subsequent work has extended GR to industrial settings, addressing practical challenges in inference efficiency or large-scale deployment~\cite{kdd26PinterestGR, Kuaishou2025OneRec, www/HeH26/PLUM, arxiv25layout}.
More recently, instruction tuning has enabled GR models to follow natural-language goals~\cite{kdd26neo_spotify, kdd26glide_spotify}, although controllable GR remains at an early stage.

\begin{figure*}[t]
\centering
\includegraphics[width=1.0\linewidth]{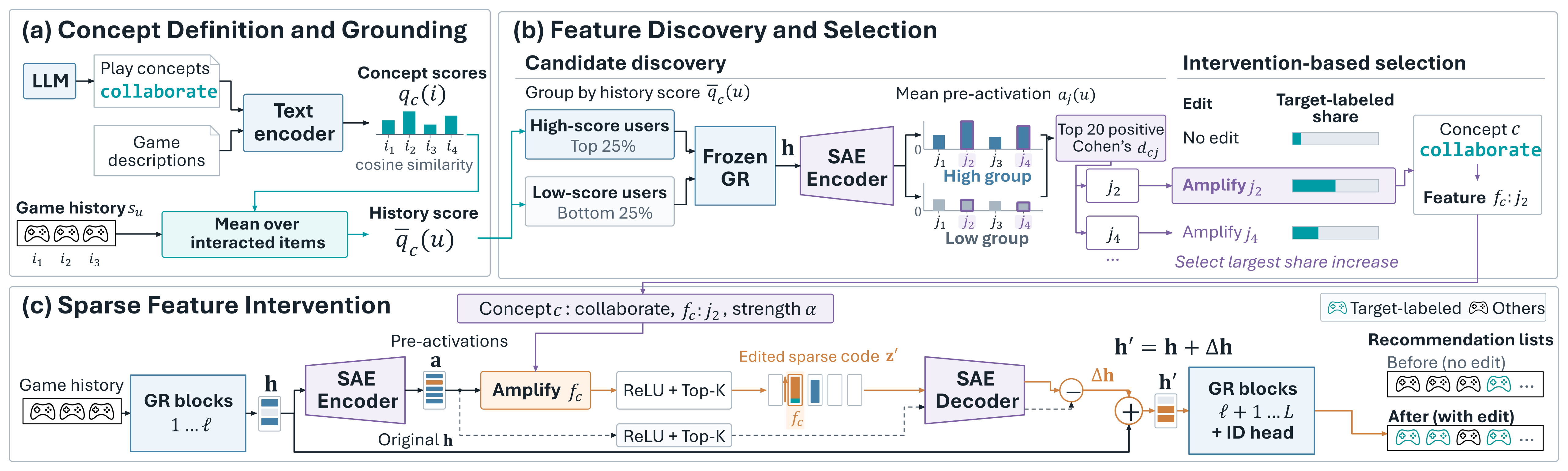}
\vspace{-5.5mm}
\caption{Overall architecture of \ours{}. We first train a GR model, freeze its parameters, and train an SAE on its hidden states. (a) Play concepts are defined and scored over games and users. (b) Features are discovered by concept association and selected by steering effect. (c) The selected feature is amplified or suppressed at inference.}\label{fig:overall_architecture}
\vspace{-4.5mm}
\end{figure*}

\subsection{Sparse Autoencoders (SAEs)}
SAEs reconstruct hidden activations through a sparsely activated dictionary~\cite{ng2011sparse, iclr24sae}, providing a basis for analyzing internal features of LLMs~\cite{iclr25saeOpenAI}.
Since each feature can be edited at inference, SAEs are widely used to steer LLMs for various downstream tasks~\cite{icml25AxBench}.
Recently, SAEs have been applied to recommendation for interpretation and control~\cite{corr26KnotsKnobs, corr25SeqRecSAE, cikm25SaulRec}.
A key benefit of SAEs is that they learn a dictionary of features from unlabeled activations, and the features can serve as steering candidates for new concepts without retraining~\cite{emnlp25SAEsurvey}.
In this work, we bring SAE steering to GR by connecting concepts to sparse features.
See Appendix~\ref{app:additional_related_work} for further discussion.

\section{Proposed Method: SteerGR}\label{sec:method}

We present \ours{}, a framework that steers GR models toward play concepts.
\ours{} builds on two models: a GR model trained for next-item prediction and then frozen, and an SAE subsequently trained on its hidden states (\S~\ref{sec:method_preliminaries}).
It then proceeds through three stages, as depicted in Figure~\ref{fig:overall_architecture}. \textit{Concept Definition and Grounding} scores games and users for each concept (\S~\ref{sec:concept_grounding}). \textit{Feature Discovery and Selection} links each concept to the sparse feature that most shifts recommendations toward it (\S~\ref{sec:feature_discovery}).
\textit{Sparse Feature Intervention} amplifies or suppresses the feature during inference (\S~\ref{sec:sparse_intervention}).

\subsection{Preliminaries}\label{sec:method_preliminaries}
\paragraph{Task Formulation.}
Let $\mathcal{U}$ and $\mathcal{I}$ denote the sets of users and items, respectively. For each user $u\in \mathcal{U}$, we represent the interaction history as a chronological item sequence $s_u=(i_1,\ldots,i_{|s_u|})$, where $i_t\in\mathcal{I}$ is the item at the $t$-th position and $|s_u|$ indicates the number of items in the sequence $s_u$. The goal is to generate the identifier of the next item $i_{|s_u|+1}$ that the user is most likely to interact with. For each play concept $c \in \mathcal C$, we aim to control the fraction of items in the recommended list labeled with the concept $c$, by intervening on internal states while keeping the recommender fixed.

\paragraph{Sparse Autoencoders.}
To steer model behaviors based on sparse features, we adopt a Top-K SAE~\citep{iclr25saeOpenAI} on hidden states of the frozen recommender. 
The SAE learns an overcomplete dictionary of features with sparse activations.
For a Transformer block $\ell$, we denote the hidden representation at input position $p$ as $\mathbf{h}_p^{(\ell)}\in\mathbb{R}^{d}$, where $d$ is the hidden dimension.
Each representation $\mathbf{h}_p^{(\ell)}$ serves as an input to the SAE.
In the following, we omit $\ell$ and $p$ for simplicity.

We first standardize $\mathbf{h}$ with the mean $\boldsymbol{\mu}_h\in\mathbb{R}^{d}$ and standard deviation $\boldsymbol{\sigma}_h\in\mathbb{R}^{d}$ of the training states.
The encoder then maps the standardized state to $m$ features ($m\gg d$) using a weight matrix $\mathbf{W}_{\mathrm{enc}}\in\mathbb{R}^{m\times d}$ and a bias term $\mathbf{b}_{\mathrm{enc}}\in\mathbb{R}^{m}$, followed by ReLU and Top-K to enforce sparsity:
\begin{equation}
\begin{aligned}
    \mathbf{x} &= (\mathbf{h}-\boldsymbol{\mu}_h)\oslash\boldsymbol{\sigma}_h,\\
    \mathbf{a} &= \mathbf{W}_{\mathrm{enc}}\mathbf{x}+\mathbf{b}_{\mathrm{enc}},\\
    \mathbf{z} &= \operatorname{TopK}_{k}(\operatorname{ReLU}(\mathbf{a})),
\end{aligned}
\label{eq:sae_encode}
\end{equation}
where $\oslash$ denotes element-wise division, $\mathbf{a}\in\mathbb{R}^{m}$ contains encoder pre-activations, and the sparse code $\mathbf{z}\in\mathbb{R}^{m}$ retains at most $k$ positive entries.

The decoder maps $\mathbf{z}$ back to the hidden space using a weight matrix $\bar{\mathbf{W}}_{\mathrm{dec}}\in\mathbb{R}^{d\times m}$ with unit-norm columns and a bias term $\mathbf{b}_{\mathrm{dec}}\in\mathbb{R}^{d}$:
\begin{equation}
    \hat{\mathbf{h}}(\mathbf{z})=\boldsymbol{\mu}_h+\boldsymbol{\sigma}_h\odot
    (\bar{\mathbf{W}}_{\mathrm{dec}}\mathbf{z}+\mathbf{b}_{\mathrm{dec}}),
\label{eq:sae_decode}
\end{equation}
where $\odot$ denotes element-wise multiplication.
We refer to each coordinate $j$, with pre-activation $a_j$, activation $z_j$, and the $j$-th decoder column as a feature.
Given $N$ training states, we train the SAE by minimizing reconstruction error:
\begin{equation}
    \mathcal{L}_{\mathrm{SAE}}=\frac{1}{Nd}\sum_{n=1}^{N}
    \left\|\mathbf{x}_n-
    (\bar{\mathbf{W}}_{\mathrm{dec}}\mathbf{z}_n+\mathbf{b}_{\mathrm{dec}})
    \right\|_2^2.
\label{eq:sae_training}
\end{equation}
Top-K enforces sparsity without an additional penalty.
Feature discovery uses the pre-activations $\mathbf{a}$, while intervention changes the sparse code $\mathbf{z}$.

\subsection{Concept Definition and Grounding}\label{sec:concept_grounding}

\paragraph{Concept Definition.}
To connect a desired goal to internal features, we first express the goal as a game-play characteristic tailored to the domain. We prompt an LLM with a domain description and its genre taxonomy to generate a fixed set $\mathcal C$ of play concepts. Each concept $c\in\mathcal C$ has a natural-language definition of the activities (\eg, \textit{collaborate}: `\textit{Players combine their efforts, skills, or roles to achieve a shared objective or complete group tasks.}'). These concepts express playing goals that span across the predefined genres in metadata\footnote{The full prompt and concepts are listed in Appendix~\ref{app:concept_definitions}.}.

\paragraph{Concept Grounding.}
We embed concept definitions and game descriptions with a shared text encoder, without manual labels for individual games.
Let $\mathbf{e}_i, \mathbf{v}_c \in \mathbb{R}^{d_{\mathrm{text}}}$ denote the unit-normalized embeddings of game $i$ and concept $c$, where $d_{\mathrm{text}}$ is the text embedding dimension.

To account for differences in cosine similarity distributions across concepts, we define the concept score as the standardized cosine similarity: 
\begin{equation}
    q_c(i)=\frac{\mathbf{e}_i^\top\mathbf{v}_c-\mu_c}{\tau_c},
\label{eq:item_concept_score}
\end{equation}
where $\mu_c$ and $\tau_c$ are the mean and standard deviation of $\mathbf{e}_i^\top\mathbf{v}_c$ over games. 
We finally assign each game to the concept with the highest score\footnote{LLM-based validation of the scores is in Appendix~\ref{app:concept_validity}.}.

For each user $u$, we compute a concept score by averaging the scores of interacted games:
\begin{equation}
    \bar{q}_c(u)=\frac{1}{|s_u|}\sum_{t=1}^{|s_u|}q_c(i_t).
\label{eq:history_concept_score}
\end{equation}
This score summarizes how prevalent concept $c$ is in a user's recent play and serves as the grouping criterion for feature discovery.

\subsection{Feature Discovery and Selection}\label{sec:feature_discovery}

\paragraph{Candidate Discovery.}

We identify candidate features that associate more strongly to each concept. 
Let $a_j(u)$ denote the pre-activation of a feature $j$ at the prediction position of user $u$'s history $s_u$.
For concept $c$, user groups with the highest and lowest quartiles of $\bar{q}_c(u)$ form a high-scoring group $\mathcal{U}_c^{+}$ and a low-scoring group $\mathcal{U}_c^{-}$.
The two groups provide a clear contrast in concept prevalence.

We compare the mean pre-activation of each feature between the groups using Cohen's $d$~\cite{routledge13cohensd}, following the prior work~\cite{www26PopSteer} as follows:
\begin{equation}
    d_{cj}=\frac{\bar a_{cj}^{+}-\bar a_{cj}^{-}}{\sigma_{cj}},
\label{eq:feature_contrast}
\end{equation}
where $\bar a_{cj}^{+}$ and $\bar a_{cj}^{-}$ are the means of $a_j(u)$ over users $u\in\mathcal{U}_c^{+}$ and $u\in\mathcal{U}_c^{-}$, respectively, and $\sigma_{cj}$ is their pooled sample standard deviation.
The standardized mean difference allows comparison across features with different scales.
Intuitively, a larger positive $d_{cj}$ indicates a stronger association between feature $j$ and concept $c$.
For each concept $c$, $N_{\mathrm{cand}}$ features with the largest positive $d_{cj}$ become candidates, and their union $U$ forms a shared candidate pool for all concepts.

\paragraph{Intervention-based Selection.}
We then select one feature for controlling each concept based on its effect on recommendations from $U$. We use this criterion since the association of feature with a concept does not guarantee that amplifying it will increase the target share~\cite{emnlp25SAERightFeatures, acl26SAE-steering, icml26corrsteer}.
On the validation split, we amplify each candidate $j\in U$ and measure the mean increase in the fraction of recommended items labeled with concept $c$, denoted as $\Delta E_{cj}$. Finally, we select the feature as:
\begin{equation}
    f_c=\argmax_{j\in U}\Delta E_{cj}.
\label{eq:feature_selection}
\end{equation}
The selected feature $f_c$ is used for amplification and suppression. This procedure successfully links each concept to the feature that most increases its target share, as shown in Table~\ref{tab:concept_feature_selection}.

\subsection{Sparse Feature Intervention}\label{sec:sparse_intervention}
We steer recommendations by editing the selected feature $f_c$ in the sparse code.
For concept $c$, the edit size is $\delta_c=\alpha\sigma_{f_c}$, where $\alpha$ controls intervention strength and $\sigma_{f_c}$ is the standard deviation of $a_{f_c}$ over the training states, following variance-based calibration~\cite{www26PopSteer}. We construct the edited code as:
\begin{equation}
    \mathbf{z}'=\begin{cases}
        \mathcal{T}(\mathbf{a}+\delta_c\mathbf{o}_{f_c}), & \text{amplify},\\
        \mathbf{z}-\delta_c\mathbf{o}_{f_c}, & \text{suppress},
    \end{cases}
\label{eq:sparse_code_edit}
\end{equation}
where $\mathbf{o}_{f_c}\in\mathbb{R}^{m}$ is the one-hot vector for $f_c$, and $\mathcal{T}(\cdot)=\operatorname{TopK}_{k}(\operatorname{ReLU}(\cdot))$.

Here, amplification raises $a_{f_c}$ and reapplies $\mathcal{T}(\cdot)$, so $\mathbf{z}'$ still retains at most $k$ positive entries.
Suppression does not reuse this rule, since the ReLU would stop $z_{f_c}$ at zero and leave positions where the feature is inactive unchanged.
Instead, we subtract $\delta_c$ directly from $z_{f_c}$, which can then become negative, similar to clamping a feature to a negative value~\cite{corr26saeclaude3sonnet}.
We add the decoded change to the original hidden state as:
\begin{equation}
\begin{aligned}
    \mathbf{h}' &= \mathbf{h}+\hat{\mathbf{h}}(\mathbf{z}')-\hat{\mathbf{h}}(\mathbf{z})\\
       &= \mathbf{h}+\boldsymbol{\sigma}_h\odot\bigl[\bar{\mathbf{W}}_{\mathrm{dec}}(\mathbf{z}'-\mathbf{z})\bigr],
\end{aligned}
\label{eq:residual_update}
\end{equation}
which preserves the reconstruction error of the original state~\cite{corr26saeclaude3sonnet}.
The recommender continues from the edited states $\mathbf{h}'$.

\section{Experimental Setup}
\label{sec:setup}

\subsection{Research Questions}\label{sec:rqs}
We address the following research questions:
\begin{itemize}[leftmargin=*,itemsep=2pt,topsep=3pt]
    \item \textbf{RQ1: Concept--feature mapping.} Are concepts effectively mapped to features?
    \item \textbf{RQ2: Controllability.} Does feature intervention successfully shift the exposure of concept?
    \item \textbf{RQ3: Control--accuracy trade-off.} Does steering improve accuracy for users whose next game matches the goal, and at what overall cost?
\end{itemize}

\subsection{Experimental Settings}\label{sec:experimental_settings}

\paragraph{Datasets.}\label{sec:datasets}
We train the GR models on proprietary interaction logs collected from a large-scale gaming platform in 2026. Each snapshot contains chronologically ordered sequence of games played by users.
For SAE, we use daily snapshots from August 2026, collected the day after training GR models.
Users are split into SAE training, validation, feature discovery, and final evaluation sets with proportions of 50/10/20/20\%, yielding 7K users in the final evaluation set. Details are in Appendix~\ref{app:setup_data}.

\noindent\textbf{Models.}
We validate \ours{} with two different models:
(i) \textbf{Atomic ID}~\cite{nips22dsi} is the recommender deployed for game discovery on the platform. It generates one token per game over a catalog of 40K games and is retrained from the previous checkpoint.
(ii) \textbf{Semantic ID}~\cite{nips23tiger} is a variant trained in the same pipeline. It represents each game using codes obtained by quantizing multimodal game embeddings with RQ-VAE~\cite{cvpr22rqvae}. It has four levels and 256 codes per level.
Both are six-layer decoder-only Transformers with a hidden size of 384, trained by next-token prediction.
For each frozen model, we train a Top-K SAE on the output of the third Transformer block. Details of the GR models and SAEs are in Appendices~\ref{app:setup_model} and~\ref{app:sae_performance}.

\noindent\textbf{Implementation Details and Metrics.}
For each SAE, we tune expansion factors $m/d\in\{8,16\}$ and $k\in\{8,16,32,64\}$.
We test amplification and suppression at $\alpha\in\{1,2,3,4,8\}$.
All interventions edit every input position. 
For controllability, we measure the \textbf{target concept share}, \ie, the target-labeled share of the top-10.
For accuracy, we adopt Recall@10 (R@10) and NDCG@10 (N@10) over \textbf{all users} and over \textbf{concept-matched users}, whose observed next-game concept matches the intervention goal. 
Further details are in Appendix~\ref{app:setup_implement}.

\begin{table}[t]
\centering
\footnotesize
\setlength{\tabcolsep}{4pt}
\renewcommand{\arraystretch}{0.94}
\begin{tabular}{@{}lrcc@{}}
\toprule
\multirow{2}{*}[-3pt]{Feature selection} & \multirow{2}{*}[-3pt]{Cohen's $d$} & \multicolumn{2}{c}{Target concept share (\%)} \\
\cmidrule(l){3-4}
& & Amplify ($\uparrow$) & Suppress ($\downarrow$) \\
\midrule
\multicolumn{4}{@{}l}{\textit{Atomic ID}} \\
\midrule
Original & -- & 5.75 & 5.75 \\
Random & $-0.03$ & 5.76 & 5.74 \\
Top-$d$ & \textbf{0.55} & 6.59 & 5.02 \\
\rowcolor{black!7} SteerGR & 0.38 & \textbf{7.95} & \textbf{4.27} \\
\midrule
\multicolumn{4}{@{}l}{\textit{Semantic ID}} \\
\midrule
Original & -- & 6.70 & 6.70 \\
Random & $-0.05$ & 6.52 & 6.85 \\
Top-$d$ & \textbf{0.85} & 6.95 & 6.55 \\
\rowcolor{black!7} SteerGR & 0.36 & \textbf{7.46} & \textbf{6.14} \\
\bottomrule
\end{tabular}
\caption{Effect of feature selection on concept association (Cohen's $d$) and target concept share. `Original' denotes no intervention, and `Top-$d$' denotes an ablation without intervention-based selection.}
\vspace{-4.5mm}
\label{tab:concept_feature_selection}
\end{table}

\begin{table*}[t]
\centering
\scriptsize\hyphenpenalty=10000\exhyphenpenalty=10000
\setlength{\tabcolsep}{2pt}
\renewcommand{\arraystretch}{1.1}
\begin{tabular}{@{}c>{\centering\arraybackslash}m{0.5cm}>{\raggedright\arraybackslash}m{2.65cm}>{\centering\arraybackslash}m{0.5cm}>{\raggedright\arraybackslash}m{2.65cm}>{\centering\arraybackslash}m{0.5cm}>{\raggedright\arraybackslash}m{2.65cm}>{\centering\arraybackslash}m{0.7cm}>{\centering\arraybackslash}m{0.5cm}>{\raggedright\arraybackslash}m{2.65cm}>{\centering\arraybackslash}m{0.7cm}@{}}
\toprule
 & \multicolumn{2}{c}{Recent plays (oldest $\rightarrow$ newest)} & \multicolumn{2}{c}{Original top-5} & \multicolumn{3}{>{\columncolor{goalAcolor}[2pt][0pt]}c}{Amplifying \textit{collaborate with others}} & \multicolumn{3}{>{\columncolor{goalBcolor}[0pt][0pt]}c@{}}{Amplifying \textit{defend against threats}} \\
\cmidrule(lr){2-3}\cmidrule(lr){4-5}\cmidrule(lr){6-8}\cmidrule(l){9-11}
 & & Game, concept & & Game, concept & & Game, concept & Before & & Game, concept & Before \\
\midrule
1 & {\normalsize\faShield*} & Base-building survival\newline \textbf{defend} & {\normalsize\faChessKnight} & Rhythm platformer\newline \textbf{tactical} & \cellcolor{goalAcolor}{\normalsize\faHandsHelping} & \cellcolor{goalAcolor}\textbf{Two-player co-op course}\newline \textbf{collaborate} & \cellcolor{goalAcolor}>10 & {\normalsize\faGem} & Hidden-threat survival\newline \textbf{collect} & 2  \\
2 & {\normalsize\faShield*} & Territory-capture game\newline \textbf{defend} & {\normalsize\faGem} & Hidden-threat survival\newline \textbf{collect} & {\normalsize\faChessKnight} & Rhythm platformer\newline \textbf{tactical} & 1 & {\normalsize\faChessKnight} & Rhythm platformer\newline \textbf{tactical} & 1  \\
3 & {\normalsize\faPuzzlePiece} & Stealth escape\newline \textbf{solve} & \cellcolor{yellow!45}{\normalsize\faTrophy} & \cellcolor{yellow!45}Battle royale brawler\newline \textbf{compete} & {\normalsize\faGem} & Hidden-threat survival\newline \textbf{collect} & 2 & \cellcolor{goalBcolor}{\normalsize\faShield*} & \cellcolor{goalBcolor}\textbf{Army battle game}\newline \textbf{defend} & \cellcolor{goalBcolor}6  \\
4 & {\normalsize\faShield*} & Wave-based shooter\newline \textbf{defend} & {\normalsize\faGem} & Morph-collect adventure\newline \textbf{collect} & {\normalsize\faGem} & Morph-collect adventure\newline \textbf{collect} & 4 & \cellcolor{yellow!45}{\normalsize\faTrophy} & \cellcolor{yellow!45}Battle royale brawler\newline \textbf{compete} & \cellcolor{yellow!45}3  \\
5 & {\normalsize\faChessKnight} & Melee PvP arena\newline \textbf{tactical} & {\normalsize\faTheaterMasks} & Character-select brawler\newline \textbf{roleplay} & \cellcolor{yellow!45}{\normalsize\faTrophy} & \cellcolor{yellow!45}Battle royale brawler\newline \textbf{compete} & \cellcolor{yellow!45}3 & \cellcolor{goalBcolor}{\normalsize\faShield*} & \cellcolor{goalBcolor}\textbf{Zombie-wave shooter}\newline \textbf{defend} & \cellcolor{goalBcolor}>10  \\
\bottomrule
\end{tabular}

\caption{Case study of amplifying two concepts for one user (features 2639 for \goalhl{goalAcolor}{\textit{collaborate}} and 1252 for \goalhl{goalBcolor}{\textit{defend}}). `Before' denotes the rank of each game in the original top-10, and yellow shading marks the ground truth game. Icons follow the concepts. Game titles and thumbnails are anonymized. Additional examples are in Appendix~\ref{app:qualitative_case}.}
\vspace{-3.5mm}
\label{tab:goal_contrast}
\end{table*}

\begin{figure}[t]
\centering
\includegraphics[width=\linewidth]{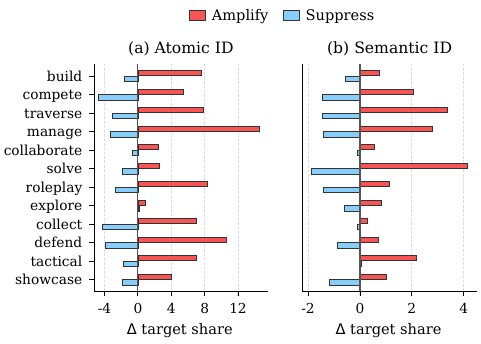}
\vspace{-4.5mm}
\caption{Change in target concept share under amplification and suppression ($\alpha=8$). Values are percentage points relative to the original model, and positive values mean more games of that concept in the top-10.}\label{fig:atomic_concept_control}
\vspace{-2.5mm}
\end{figure}

\section{Experimental Results}
\label{sec:results}
\begin{figure}[t]
\centering
\includegraphics[width=\linewidth]{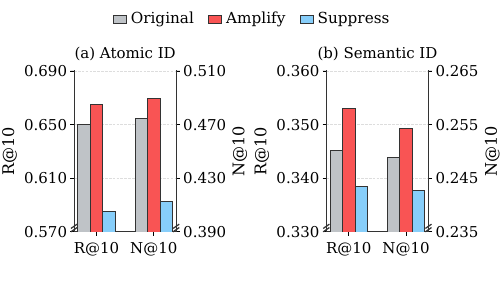}
\vspace{-9mm}
\caption{Accuracy for concept-matched users under intervention, averaged over all concepts.}
\vspace{-4.5mm}
\label{fig:atomic_next_game_relevance}
\end{figure}

\subsection{Main Results}
\label{sec:main_results}
\paragraph{RQ1: Concept--Feature Mapping.}
Table~\ref{tab:concept_feature_selection} illustrates the effectiveness of our feature selection by comparing it with two selection baselines on held-out users. 
Top-$d$ takes the candidate with the highest Cohen's $d$ and Random uses a feature of similar activation frequency.
\textbf{(i) Intervention-based selection identifies effective steering features that generalize to held-out users.}
Under amplification, \ours{} increases the target concept share by 38.4\% and 11.5\% for Atomic ID and Semantic ID, respectively.
These gains are 2.6$\times$ and 3.0$\times$ those of Top-$d$.
Under suppression, it reduces the share by 25.6\% and 8.3\%, whereas Random changes it by less than 3\%. It confirms that the steering effects do not stem from arbitrary feature edits.
\textbf{(ii) A strong concept association does not guarantee a strong steering effect.}
Top-$d$ features are the most strongly associated with their concepts, yet they change exposure less than \ours{}.
This exhibits that features for steering should be selected by their steering effects rather than by association alone, which is consistent with existing findings in LLMs~\cite{emnlp25SAERightFeatures, acl26SAE-steering}\footnote{Further analysis of the features is in Appendix~\ref{app:feature}.}.

\paragraph{RQ2: Controllability.}
Figure~\ref{fig:atomic_concept_control} shows that the target concept share moves in the intended direction under both amplification and suppression.
Amplification increases the target concept share for all twelve concepts in both models, by 113.6\% for Atomic ID and 24.6\% for Semantic ID on average.
Suppression reduces it by 42.5\% and 13.7\%, respectively, with decreases for all concepts except \textit{explore} in Atomic ID and \textit{tactical} in Semantic ID.
The magnitude varies across concepts, with amplification gains ranging from 35\% (\textit{explore}) to 224\% (\textit{build}) for Atomic ID and from 4\% (\textit{collect}) to 64\% (\textit{solve}) for Semantic ID.
This demonstrates that sparse feature intervention can steer the concept exposure in recommendation results, although the strength of control depends on the concept\footnote{More features per concept steer further but also change other concepts more. Details are provided in Appendix~\ref{app:feature_multi}.}.

\begin{figure}[t]
\centering
\includegraphics[width=\linewidth]{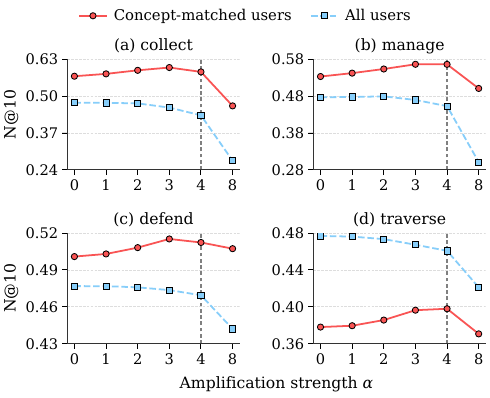}
\vspace{-4.5mm}
\caption{Accuracy of concept-matched users and all users under amplification across $\alpha$. We show the four concepts of the Atomic ID with the largest gains in target concept share on the validation split.}
\label{fig:atomic_control_tradeoff}
\vspace{-3.5mm}
\end{figure}

\paragraph{RQ3: Control--Accuracy Trade-off.}
Figure~\ref{fig:atomic_next_game_relevance} presents accuracy of concept-matched users, whose next game belongs to the target concept. Here, effective steering should \textit{raise the accuracy under amplification} and \textit{lower it under suppression}. 
Amplification increases R@10 by 2.4\% and 2.3\% for Atomic ID and Semantic ID, whereas suppression reduces it by 10.0\% and 1.9\%.
This suggests that selected features successfully capture the preferences of users who seek the concept.
We further examine how stronger interventions affect accuracy over all users. 
Figure~\ref{fig:atomic_control_tradeoff} shows accuracy across intervention strengths for the four concepts of the Atomic ID model.
Concept-matched accuracy peaks at $\alpha=3$ or $4$ with gains of 2.9--6.2\%, and overall accuracy drops by 7.3--42.9\% at $\alpha=8$. This indicates that strong interventions overly push concept-matched games into results, so strength should be balanced against overall accuracy.

\subsection{Case Study}
\label{sec:case_study}
Table~\ref{tab:goal_contrast} shows a case study in which we amplify two different concepts for the user.
Amplifying \textit{collaborate with others} places a collaboration-labeled game first, whereas amplifying \textit{defend against threats} brings two defense-labeled games into the list.
Notably, the observed next game stays in the top five, illustrating that steering can shift recommendations toward the requested goal while preserving the user's preference.

\section{Conclusion}

We present \ours{}, an SAE-based framework that steers generative recommendation toward play concepts.
\ours{} links each concept to a sparse feature by its effect on recommendations and controls the feature while keeping the recommender frozen.
Extensive experiments on large-scale industrial game datasets show that \ours{} successfully finds effective steering features, and intervening on them shifts concept exposure in the intended direction.
Since the recommender remains frozen, \ours{} can serve diverse user requests or platform objectives without retraining. 
We believe that our work bridges SAE steering and generative recommendation, demonstrating the potential for controlling recommendations toward explicit goals. 

\section{Limitations}\label{sec:limitation}
Our work has two limitations that suggest directions for future work.
(i) We evaluate \ours{} on twelve LLM-generated play concepts grounded in game descriptions.
However, evolving needs of platforms and discovery goals of users may require diverse steering criteria (\eg, popularity and novelty).
Future work could extend \ours{} with controls for more diverse criteria.
(ii) We demonstrate that a Top-K SAE~\cite{iclr25saeOpenAI} can support concept steering in generative recommendation as a proof-of-concept.
However, recent works on SAEs have explored a broader range of architectures and training strategies~\cite{nips24end2endsae, icml25MatryoshkaSAE, iclr26TSAE}.
Future work could adapt these advances to generative recommendation to investigate more flexible concept control and assess their effects on recommendation quality.

\section*{Ethics Statement}

This work fully complies with the ACL Ethics Policy. We declare that there are no ethical issues in this paper. The scientific artifacts we have utilized are publicly available for research under permissive licenses, and the utilization of these tools is consistent with their intended applications.

\bibliography{references}

\appendix

\newpage

\section{Additional Related Work}\label{app:additional_related_work}
\begin{table}[t]
\centering
\small
\begin{tabular}{lrr}
\toprule
 & Atomic ID & Semantic ID \\
\midrule
Catalog size (games) & 40K & 100K \\
SAE training states & 79K & 381K \\
Validation users & 3.2K & 3.5K \\
Discovery users & 6.4K & 7.0K \\
Evaluation users & 6.3K & 6.9K \\
Mean history length & 616 & 213 \\
\bottomrule
\end{tabular}
\caption{Data statistics used for the SAE study.}\label{tab:data_statistics}
\vspace{-4.5mm}
\end{table}

\paragraph{SAEs for Interpretation and Steering.}
Subsequent work on SAEs~\cite{iclr24sae} scales them to larger LLMs~\cite{iclr25saeOpenAI, corr26saeclaude3sonnet}, with improved activations~\cite{corr24saejumprelu} and open SAEs for public models~\cite{corr24GemmaScope}.
Beyond text, SAEs also yield interpretable features in vision--language models~\cite{nips25SAE-VLM}. 
To evaluate SAEs beyond reconstruction, recent benchmarks measure the interpretability of features in downstream tasks~\cite{icml25SAEBench}, as well as consistency across training runs~\cite{acl26PW-MCC}.
Editing these features can steer model behavior across tasks, from classification to reasoning~\cite{iclr25shift, acl26SAE-steering}.
However, which features to edit remains a key question: a feature that responds to an input pattern does not always change the output, and choosing features by their effect on the output improves steering~\cite{emnlp25SAERightFeatures, acl26SAE-steering, icml26corrsteer}.
Following this view, we find candidates by their association with user histories and select among them by their effect on recommendations.

\paragraph{SAEs in Recommendation.}
In recommendation, SAE features have been linked to semantic concepts and user intents across recommender architectures~\cite{tois26RecSAE, corr26SAERec}, and a prediction-aware objective keeps the extracted concepts consistent with the model's predictions~\cite{AAAI26Monosemanticity4Rec}.
For control, prior work steers sequential recommenders with features tied to item attributes~\cite{corr25SeqRecSAE} and reduces popularity bias with features identified from synthetic histories~\cite{www26PopSteer}.
While these studies interpret features or control predefined attributes and popularity, we define play concepts in text and ground them in game descriptions.
We then evaluate steering in fixed GR models by how it shifts recommendations toward each concept and how it affects next-game relevance.

\paragraph{Generative Recommendation.}
In line with generative retrieval~\cite{nips22dsi, tois25genirsurvey, emnlp23GLEN}, GR formulates recommendation as generating the identifier of the next item~\cite{Geng0FGZ22P5, HuaXGZ23P5Howtoindex}.
Most studies focus on constructing item identifiers, either as semantic IDs quantized from item embeddings~\cite{nips23tiger, ZhengHLCZCW24LCRec, wang2024letter, iclr26catalogllm} or as textual IDs derived from item metadata~\cite{Tan24IDGenRec, corr26agentictagger, acl25GRAM, recsys26bonsai, emnlp25grut}.
GR has also been adopted in industrial systems~\cite{corr26netflixGR, sigir26genrec_jd, corr26tabao_gr, acl26AuthGR, recsys26to-rope}, where a single generative model replaces multi-stage pipelines~\cite{Kuaishou2025OneRec, acl26OneRecThink, kdd26PinterestGR}.
They improve how GR represents items and learns user preferences, whereas we steer the outputs of a trained GR model through its sparse features without retraining.

\begin{table}[t]
\centering
\small
\begin{tabular}{lcc}
\toprule
Hyperparameter & Atomic ID & Semantic ID \\
\midrule
\multicolumn{3}{l}{\textit{GR model}} \\
\midrule
Optimizer & AdamW & AdamW \\
Learning rate & $10^{-4}$ & $10^{-4}$ \\
Adam $(\beta_1,\beta_2)$ & (0.95, 0.99) & (0.95, 0.99) \\
Weight decay & 0.06 & 0.06 \\
\midrule
\multicolumn{3}{l}{\textit{SAE}} \\
\midrule
Dictionary size $m$ & 3,072 & 3,072 \\
Active features $k$ & 32 & 64 \\
Optimizer & Adam & Adam \\
Learning rate & $10^{-3}$ & $10^{-3}$ \\
Batch size & 2,048 & 2,048 \\
\midrule
\multicolumn{3}{l}{\textit{Concept discovery and selection}} \\
\midrule
Text embedding $d_{\mathrm{text}}$ & 3,072 & 3,072 \\
Candidates $N_{\mathrm{cand}}$ & 20 & 20 \\
Candidate pool $|U|$ & 188 & 169 \\
Selection strength $\alpha$ & 4 & 4 \\
\bottomrule
\end{tabular}
\caption{Final hyperparameters for \ours{}.}\label{tab:hyperparams}
\vspace{-4.5mm}
\end{table}

\section{Additional Experimental Setup}\label{app:setup}
\subsection{Datasets}\label{app:setup_data}

Table~\ref{tab:data_statistics} summarizes the data statistics for the SAE study.
Both models are trained on snapshots from mid-August 2026, and the four splits contain disjoint users.
Semantic ID histories are shorter as each game takes four tokens of the same 1,024-token context.
Atomic ID assigns a dedicated token to each of the games, whereas Semantic ID composes shared codes and thus covers a larger catalog. 

\begin{table*}[t]
\centering
\footnotesize
\setlength{\tabcolsep}{3pt}
\renewcommand{\arraystretch}{1.1}
\begin{tabular}{p{0.115\textwidth}p{0.245\textwidth}p{0.585\textwidth}}
\toprule
Short label & Label & Definition \\
\midrule
build & build and customize & Players place, modify, or decorate objects and structures to create a personalized environment or unique character appearance. \\
compete & compete for dominance & Players engage in structured matches or challenges to outperform rivals and secure a higher standing on a leaderboard. \\
traverse & traverse challenging terrain & Players navigate complex environments by jumping, climbing, and dodging obstacles to reach a specific destination. \\
manage & manage resources and wealth & Players harvest, trade, or invest in-game assets to grow their economy and unlock better equipment or upgrades. \\
collaborate & collaborate with others & Players combine their efforts, skills, or roles to achieve a shared objective or complete group tasks. \\
solve & solve logic puzzles & Players decipher patterns, analyze clues, or rearrange objects to progress through interactive brain-teasing sequences. \\
roleplay & roleplay and perform & Players act out specific personas, take on social jobs, or participate in scripted scenarios to interact with the world. \\
explore & explore hidden worlds & Players uncover new areas, search for secret lore, or chart vast environments filled with unique sights and sounds. \\
collect & collect rare items & Players hunt for, unlock, or trade for unique loot, trophies, or artifacts to display their success. \\
defend & defend against threats & Players hold defensive positions or utilize fortifications to protect themselves and their property from incoming hostile forces. \\
tactical & perform tactical maneuvers & Players utilize movement, positioning, and timing to outmaneuver opponents or navigate dangerous environments efficiently. \\
showcase & showcase personal style & Players display their creative choices, outfits, or custom builds for others to view and appreciate. \\
\bottomrule
\end{tabular}
\caption{LLM-generated play concepts. `Short labels' denotes the name used in the paper, and the definitions are embedded to compute concept scoring.}\label{tab:play_concept_definitions}
\end{table*}

\subsection{Models}\label{app:setup_model}
GR models use six attention heads and a maximum context of 1,024 tokens, and are trained to minimize the cross-entropy of each ID token given the preceding inputs:
\begin{equation}
    \mathcal{L}_{\mathrm{GR}}=-\sum_{p}\log P_\theta(y_p\mid \mathcal{H}_{<p}),
\label{eq:gr_training}
\end{equation}
where $y_p$ is the $p$-th identifier token, \ie, a game token for Atomic ID or one of the four codes for Semantic ID, and $\mathcal{H}_{<p}$ denotes all preceding inputs.
Semantic ID predicts the four codes of the next game level by level, using a separate output head for each level.

The Atomic ID SAE is trained on one hidden state per next-game prediction.
The Semantic ID SAE is trained on four positions per next game under observed code prefixes: the history boundary predicting the first code and the three positions predicting the remaining codes.
These four positions share one SAE, while concept discovery uses only the first-code prediction boundary.

\subsection{Implementation Details and Metrics}\label{app:setup_implement}
Table~\ref{tab:hyperparams} lists the final hyperparameters.
Among the searched SAE configurations, we choose the smallest $k$ whose SAE explains at least 70\% of the variance, has at most 1\% dead features, and lowers Recall@300 by at most 2\% when its reconstruction replaces the hidden state.
Each SAE is early-stopped on the reconstruction error of held-out training states.

We generate twelve play concepts with \texttt{gemini-3.1-flash-lite-preview}, and embed concepts and game descriptions with \texttt{text-embedding-3-large}.
User concept scores use only the 20 most recent interactions, excluding games without a concept score.
For Atomic ID, the top-10 list ranks the logits of all games, excluding padding and unknown tokens.
For Semantic ID, we decode with a beam size of 300.
The intervention is applied only during the history pass. Games without a concept score receive no label but count toward the denominator and overall accuracy.

\begin{table}[t]
\centering
\footnotesize
\setlength{\tabcolsep}{3pt}
\renewcommand{\arraystretch}{0.94}
\begin{tabular}{@{}lcccccc@{}}
\toprule
\multirow{2}{*}{Model} & \multirow{2}{*}{$R^2$} & \multirow{2}{*}{$L_0$} & \multicolumn{3}{c}{N@10 with block output} & \multirow{2}{*}{Rec.} \\
\cmidrule(lr){4-6}
& & & Original & SAE & Mean & \\
\midrule
Atomic & 0.750 & 32 & 0.477 & 0.408 & 0.064 & 83.4\% \\
Semantic & 0.951 & 64 & 0.214 & 0.206 & 0.036 & 95.5\% \\
\bottomrule
\end{tabular}
\caption{SAE quality on final evaluation users. `Rec.' denotes the recovered share, computed as $(\mathrm{SAE}-\mathrm{Mean})/(\mathrm{Original}-\mathrm{Mean})$.}\label{tab:sae_performance}
\end{table}

\begin{table}[t]
\centering
\footnotesize
\setlength{\tabcolsep}{4pt}
\renewcommand{\arraystretch}{0.94}
\begin{tabular}{@{}lrrr@{}}
\toprule
\multirow{2}{*}{Concept} & \multicolumn{2}{c}{AUC} & \multirow{2}{*}{Genres} \\
\cmidrule(lr){2-3}
& GPT & Gemini & \\
\midrule
build and customize & 0.75 & 0.96 & 10 \\
compete for dominance & 0.84 & 0.90 & 10 \\
traverse challenging terrain & 0.96 & 0.97 & 6 \\
manage resources and wealth & 0.90 & 0.96 & 9 \\
collaborate with others & 0.73 & 0.81 & 12 \\
solve logic puzzles & 0.95 & 1.00 & 5 \\
roleplay and perform & 0.84 & 0.87 & 9 \\
explore hidden worlds & 0.81 & 0.95 & 8 \\
collect rare items & 0.78 & 0.92 & 10 \\
defend against threats & 0.91 & 0.97 & 8 \\
perform tactical maneuvers & 0.83 & 0.83 & 9 \\
showcase personal style & 0.80 & 0.90 & 9 \\
\midrule
\textbf{Mean} & \textbf{0.84} & \textbf{0.92} & -- \\
\bottomrule
\end{tabular}
\caption{AUC of concept scores, with LLM judgments on 400 games. `Genres' denotes the number of genres out of 15 among the games selected for each concept.}
\label{tab:concept_validity}
\end{table}

\subsection{Details of Concept Scoring}\label{app:concept_definitions}

Table~\ref{tab:play_concept_definitions} lists all concepts and definitions.
Figure~\ref{fig:concept_prompt} shows the prompt used to generate the concepts.
We embed the definition sentence of each concept.

\section{Additional Experimental Results}\label{app:experimental_results}

\subsection{SAE Quality}\label{app:sae_performance}
Following prior SAE studies~\cite{iclr25saeOpenAI, nips25SAE-VLM, corr25SeqRecSAE}, we prove that both SAEs are well trained as shown in Table~\ref{tab:sae_performance}. 
We evaluate SAEs on final evaluation users, who are unseen in SAE training and checkpoint selection. 
With only 1--2\% of the 3,072 features active per state, they explain 75.0--95.1\% of the hidden-state variance and leave no feature dead.
The reconstructed vectors also preserve recommendation quality, lowering N@10 by 14.4\% for Atomic ID and 3.7\% for Semantic ID when replacing the original block output.
This drop is comparable to or smaller than the 7--24\% decrease that prior work reports as largely stable for SASRec~\cite{tois26RecSAE}.
It also does not affect intervention, since Eq.\ref{eq:residual_update} keeps the original state and adds only the change caused by the edit.

\begin{figure}[t]
\centering
\includegraphics[width=\linewidth]{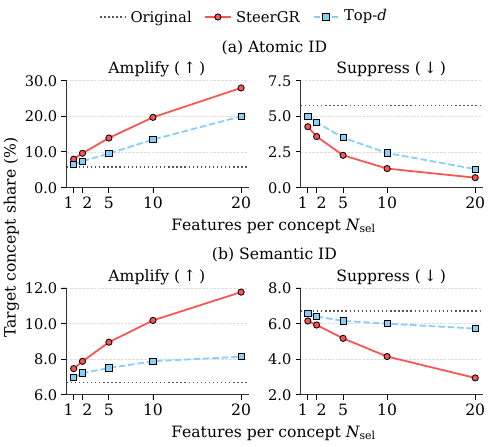}
\caption{Target concept share across the number of selected features per concept, $N_{\mathrm{sel}}\in\{1,2,5,10,20\}$ ($\alpha=4$). We use $N_{\mathrm{sel}}=1$ by default.}
\label{fig:multi_feature}
\end{figure}

\begin{figure}[t]
\centering
\includegraphics[width=\linewidth]{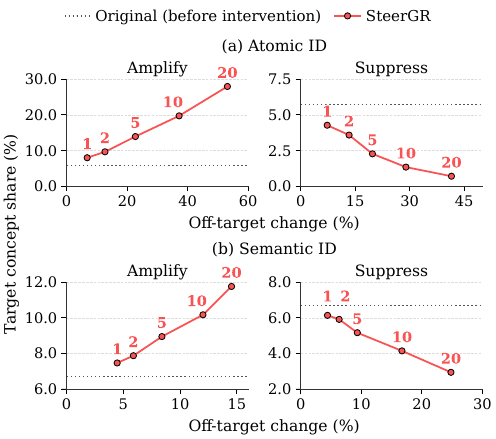}
\caption{Target concept share versus off-target change for \ours{} across the number of features per concept $N_{\mathrm{sel}}\in\{1,2,5,10,20\}$ ($\alpha=4$). Numbers mark $N_{\mathrm{sel}}$.}\label{fig:multi_feature_offtarget}
\end{figure}

\subsection{Validity of Play Concepts}
\label{app:concept_validity}
Table~\ref{tab:concept_validity} validates the concept grounding with two LLM judges that did not generate the concepts, \texttt{gpt-5.5} and \texttt{gemini-3.8-flash}.
We sample 400 games, 25 per concept and 100 uniformly.
Each judge reads definitions of all concepts and the game description, and selects every concept that clearly describes the game with supporting quotes, as shown in Figure~\ref{fig:judge_prompt}.

\textbf{(i) Concept scores agree with the judges.}
The games that the judges select for a concept receive higher concept scores than the other games, with mean AUCs of 0.84 and 0.92 and at least 0.73 for every concept.
For games assigned to $c$, the judges select $c$ in 4.1 and 8.2 times as often for other games.
\textbf{(ii) Each concept spans multiple genres.}
For each concept, the games selected by either judge span five or more of the 15 genres. 
This indicates that the concepts capture playing goals that a single genre label does not express.

\subsection{Analysis of Features}\label{app:feature}

\subsubsection{Multiple Features per Concept}\label{app:feature_multi}
\paragraph{Target Share.}
Figure~\ref{fig:multi_feature} shows that feature selection method of \ours{} remains effective when multiple features are selected per concept.
We edit the top $N_{\mathrm{sel}}$ candidates per concept with $\alpha=4$, ranked by $\Delta E_{cj}$ for \ours{} and by $d_{cj}$ for Top-$d$.
Both methods change the target share more as $N_{\mathrm{sel}}$ grows, and \ours{} changes it 1.1--3.8$\times$ more than Top-$d$ at every $N_{\mathrm{sel}}$.

\paragraph{Off-target Change.}
Figure~\ref{fig:multi_feature_offtarget} shows that this gain comes with a trade-off, as more features also change the shares of other concepts.
Relative off-target change is the mean absolute change in the shares of the other eleven concepts, expressed as a percentage of their mean share before intervention.
In both models, increasing $N_{\mathrm{sel}}$ strengthens amplification or suppression of the target concept but also increases off-target change.
We therefore retain one feature per concept in the main experiments, following prior works~\cite{corr26saeclaude3sonnet, corr25SeqRecSAE}.

\subsubsection{Swapping Features Across Concepts}\label{app:feature_swap}

\begin{figure}[t]
\centering
\includegraphics[width=\linewidth]{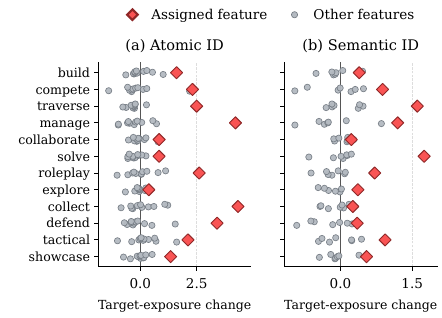}
\caption{Effects of the assigned feature on the target concept share, compared with other features. Results are for held-out users under amplification ($\alpha=4$). Changes are in percentage points relative to the original model.}
\label{fig:concept_feature_assignment}
\end{figure}

Figure~\ref{fig:concept_feature_assignment} illustrates that each selected feature steers its own concept rather than shifting recommendations toward any concept.
We measure the change in every concept share under every selected feature on final evaluation users.
The selected assignment yields a larger gain in target share than all 200K random reassignments of features to concepts in both models ($p\approx5\times10^{-6}$).

\subsubsection{Associations with Concepts}\label{app:feature_activity}
\begin{figure}[t]
\centering
\includegraphics[width=\linewidth]{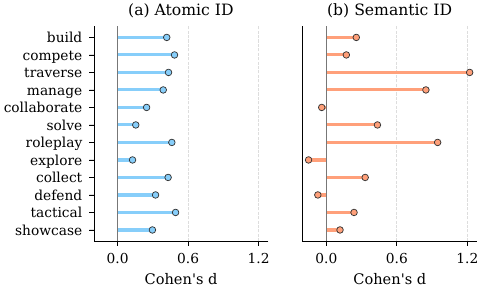}
\caption{Concept association of the selected features, measured by Cohen's $d$ of pre-activations between high- and low-scoring user groups.}\label{fig:concept_feature_associations}
\vspace{-3.5mm}

\end{figure}

Figure~\ref{fig:concept_feature_associations} shows the $d_{cj}$ of each selected feature for its assigned concept.
All twelve Atomic ID features have positive contrasts, whereas Semantic ID shows small negative contrasts for \textit{collaborate}, \textit{explore}, and \textit{defend}.
These contrasts are stable, with a correlation of 0.996 between discovery and held-out users across all 144 pairs of selected features and concepts.
As in Table~\ref{tab:concept_feature_selection}, these contrasts are smaller than those of Top-$d$ features, yet \ours{} features steer more, showing that a strong association does not guarantee effective steering.

\subsection{Additional Case Study}\label{app:qualitative_case}

\paragraph{Case 1. Amplifying different concepts.}
Tables~\ref{tab:goal_contrast_a} and~\ref{tab:goal_contrast_c} are extended examples of amplifying different concepts.
In Table~\ref{tab:goal_contrast_a}, amplifying \textit{collaborate} and \textit{defend} adds one and two matching games to the top five, while the same top three games stay in place; both \textit{defend} additions come from the user's recent plays.
In Table~\ref{tab:goal_contrast_c}, \textit{collaborate} adds one matching game and \textit{manage} adds two, while the top-ranked game stays first in both lists.

\paragraph{Case 2. Amplifying and suppressing the same concept.}
Table~\ref{tab:amp_supp} contrasts amplification and suppression of the same concept for one user.
This user shows the largest change in both directions across all twelve goals.
Amplifying \textit{traverse} fills the top five with wall-jump courses, whereas suppressing it replaces \textit{traverse} games with \textit{collaborate} and \textit{defend} games.

\paragraph{Case 3. Suppressing different concepts.}
Table~\ref{tab:supp_contrast} suppresses two concepts separately for one user at $\alpha=4$.
The original result holds two \textit{compete}- and two \textit{collect} games.
Suppressing \textit{compete} removes both \textit{compete} games and keeps both \textit{collect} games, whereas suppressing \textit{collect} removes both \textit{collect} games and leaves the \textit{compete} count unchanged.

\FloatBarrier
\begin{table*}[t]
\centering
\scriptsize\hyphenpenalty=10000\exhyphenpenalty=10000
\setlength{\tabcolsep}{2pt}
\renewcommand{\arraystretch}{1.1}
\begin{tabular}{@{}c>{\centering\arraybackslash}m{0.5cm}>{\raggedright\arraybackslash}m{2.65cm}>{\centering\arraybackslash}m{0.5cm}>{\raggedright\arraybackslash}m{2.65cm}>{\centering\arraybackslash}m{0.5cm}>{\raggedright\arraybackslash}m{2.65cm}>{\centering\arraybackslash}m{0.7cm}>{\centering\arraybackslash}m{0.5cm}>{\raggedright\arraybackslash}m{2.65cm}>{\centering\arraybackslash}m{0.7cm}@{}}
\toprule
 & \multicolumn{2}{c}{Recent plays (oldest $\rightarrow$ newest)} & \multicolumn{2}{c}{Original top-5} & \multicolumn{3}{>{\columncolor{goalAcolor}[2pt][0pt]}c}{Amplifying \textit{collaborate with others}} & \multicolumn{3}{>{\columncolor{goalBcolor}[0pt][0pt]}c@{}}{Amplifying \textit{defend against threats}} \\
\cmidrule(lr){2-3}\cmidrule(lr){4-5}\cmidrule(lr){6-8}\cmidrule(l){9-11}
 & & Game, concept & & Game, concept & & Game, concept & Before & & Game, concept & Before \\
\midrule
1 & {\normalsize\faStar} & Free-item obstacle course\newline \textbf{showcase} & {\normalsize\faTrophy} & FPS duel game\newline \textbf{compete} & {\normalsize\faTrophy} & FPS duel game\newline \textbf{compete} & 1 & {\normalsize\faTrophy} & FPS duel game\newline \textbf{compete} & 1 \\
2 & {\normalsize\faCompass} & Downhill sled runner\newline \textbf{explore} & {\normalsize\faCompass} & Downhill sled runner\newline \textbf{explore} & {\normalsize\faCompass} & Downhill sled runner\newline \textbf{explore} & 2 & {\normalsize\faCompass} & Downhill sled runner\newline \textbf{explore} & 2 \\
3 & {\normalsize\faShield*} & Bomb-passing game\newline \textbf{defend} & {\normalsize\faTrophy} & FPS deathmatch game\newline \textbf{compete} & {\normalsize\faTrophy} & FPS deathmatch game\newline \textbf{compete} & 3 & {\normalsize\faTrophy} & FPS deathmatch game\newline \textbf{compete} & 3 \\
4 & {\normalsize\faShield*} & Team arena shooter\newline \textbf{defend} & {\normalsize\faTrophy} & Racket sports game\newline \textbf{compete} & {\normalsize\faTrophy} & Racket sports game\newline \textbf{compete} & 4 & \cellcolor{goalBcolor}{\normalsize\faShield*} & \cellcolor{goalBcolor}\textbf{Bomb-passing game}\newline \textbf{defend} & \cellcolor{goalBcolor}6 \\
5 & {\normalsize\faHandsHelping} & Co-op road-trip survival\newline \textbf{collaborate} & {\normalsize\faCoins} & Open-world action RPG\newline \textbf{manage} & \cellcolor{goalAcolor}{\normalsize\faHandsHelping} & \cellcolor{goalAcolor}\textbf{Team block co-op course}\newline \textbf{collaborate} & \cellcolor{goalAcolor}9 & \cellcolor{goalBcolor}{\normalsize\faShield*} & \cellcolor{goalBcolor}\textbf{Team arena shooter}\newline \textbf{defend} & \cellcolor{goalBcolor}8 \\
\bottomrule
\end{tabular}
\caption{Case study of amplifying two concepts ($\alpha=4$; features 2639 for \goalhl{goalAcolor}{\textit{collaborate}} and 1252 for \goalhl{goalBcolor}{\textit{defend}}). `Before' denotes the rank of each game in the original top-10, and shading marks games labeled with the amplified concept. Icons follow the concepts. Game titles and thumbnails are anonymized.}
\label{tab:goal_contrast_a}
\end{table*}

\begin{table*}[t]
\centering
\scriptsize\hyphenpenalty=10000\exhyphenpenalty=10000
\setlength{\tabcolsep}{2pt}
\renewcommand{\arraystretch}{1.1}
\begin{tabular}{@{}c>{\centering\arraybackslash}m{0.5cm}>{\raggedright\arraybackslash}m{2.65cm}>{\centering\arraybackslash}m{0.5cm}>{\raggedright\arraybackslash}m{2.65cm}>{\centering\arraybackslash}m{0.5cm}>{\raggedright\arraybackslash}m{2.65cm}>{\centering\arraybackslash}m{0.7cm}>{\centering\arraybackslash}m{0.5cm}>{\raggedright\arraybackslash}m{2.65cm}>{\centering\arraybackslash}m{0.7cm}@{}}
\toprule
 & \multicolumn{2}{c}{Recent plays (oldest $\rightarrow$ newest)} & \multicolumn{2}{c}{Original top-5} & \multicolumn{3}{>{\columncolor{goalAcolor}[2pt][0pt]}c}{Amplifying \textit{collaborate with others}} & \multicolumn{3}{>{\columncolor{goalBcolor}[0pt][0pt]}c@{}}{Amplifying \textit{manage resources and wealth}} \\
\cmidrule(lr){2-3}\cmidrule(lr){4-5}\cmidrule(lr){6-8}\cmidrule(l){9-11}
 & & Game, concept & & Game, concept & & Game, concept & Before & & Game, concept & Before \\
\midrule
1 & {\normalsize\faChessKnight} & Co-op horror escape\newline \textbf{tactical} & {\normalsize\faGem} & Forest survival game\newline \textbf{collect} & {\normalsize\faGem} & Forest survival game\newline \textbf{collect} & 1 & {\normalsize\faGem} & Forest survival game\newline \textbf{collect} & 1 \\
2 & {\normalsize\faHandsHelping} & Toss-and-carry course\newline \textbf{collaborate} & {\normalsize\faGem} & Monster escape survival\newline \textbf{collect} & {\normalsize\faGem} & Monster escape survival\newline \textbf{collect} & 2 & \cellcolor{goalBcolor}{\normalsize\faCoins} & \cellcolor{goalBcolor}\textbf{Steal-and-defend tycoon}\newline \textbf{manage} & \cellcolor{goalBcolor}9 \\
3 & {\normalsize\faHandsHelping} & Piggyback climb course\newline \textbf{collaborate} & {\normalsize\faGem} & Team duel shooter\newline \textbf{collect} & {\normalsize\faGem} & Team duel shooter\newline \textbf{collect} & 3 & {\normalsize\faTrophy} & Elimination party game\newline \textbf{compete} & 5 \\
4 & {\normalsize\faHandsHelping} & Shared-plank course\newline \textbf{collaborate} & {\normalsize\faChessKnight} & Clicker simulator\newline \textbf{tactical} & {\normalsize\faTrophy} & Elimination party game\newline \textbf{compete} & 5 & \cellcolor{goalBcolor}{\normalsize\faCoins} & \cellcolor{goalBcolor}\textbf{Farming tycoon}\newline \textbf{manage} & \cellcolor{goalBcolor}6 \\
5 & {\normalsize\faTheaterMasks} & Horror morph roleplay\newline \textbf{roleplay} & {\normalsize\faTrophy} & Elimination party game\newline \textbf{compete} & \cellcolor{goalAcolor}{\normalsize\faHandsHelping} & \cellcolor{goalAcolor}\textbf{Two-player co-op course}\newline \textbf{collaborate} & \cellcolor{goalAcolor}>10 & {\normalsize\faGem} & Team duel shooter\newline \textbf{collect} & 3 \\
\bottomrule
\end{tabular}
\caption{Case study of amplifying two concepts ($\alpha=4$; features 2639 for \goalhl{goalAcolor}{\textit{collaborate}} and 2787 for \goalhl{goalBcolor}{\textit{manage}}). `Before' denotes the rank of each game in the original top-10, and shading marks games labeled with the amplified concept. Icons follow the concepts. Game titles and thumbnails are anonymized.}
\label{tab:goal_contrast_c}
\end{table*}

\begin{table*}[t]
\centering
\scriptsize\hyphenpenalty=10000\exhyphenpenalty=10000
\setlength{\tabcolsep}{2pt}
\renewcommand{\arraystretch}{1.1}
\begin{tabular}{@{}c>{\centering\arraybackslash}m{0.5cm}>{\raggedright\arraybackslash}m{2.65cm}>{\centering\arraybackslash}m{0.5cm}>{\raggedright\arraybackslash}m{2.65cm}>{\centering\arraybackslash}m{0.5cm}>{\raggedright\arraybackslash}m{2.65cm}>{\centering\arraybackslash}m{0.7cm}>{\centering\arraybackslash}m{0.5cm}>{\raggedright\arraybackslash}m{2.65cm}>{\centering\arraybackslash}m{0.7cm}@{}}
\toprule
 & \multicolumn{2}{c}{Recent plays (oldest $\rightarrow$ newest)} & \multicolumn{2}{c}{Original top-5} & \multicolumn{3}{>{\columncolor{goalAcolor}[2pt][0pt]}c}{Amplifying \textit{traverse challenging terrain}} & \multicolumn{3}{>{\columncolor{goalAcolor}[0pt][0pt]}c@{}}{Suppressing \textit{traverse challenging terrain}} \\
\cmidrule(lr){2-3}\cmidrule(lr){4-5}\cmidrule(lr){6-8}\cmidrule(l){9-11}
 & & Game, concept & & Game, concept & & Game, concept & Before & & Game, concept & Before \\
\midrule
1 &  & Driving game\newline -- & {\normalsize\faHandsHelping} & Vehicle stunt simulator\newline \textbf{collaborate} & \cellcolor{goalAcolor}{\normalsize\faRunning} & \cellcolor{goalAcolor}\textbf{Wall-jump course}\newline \textbf{traverse} & \cellcolor{goalAcolor}2 & {\normalsize\faHandsHelping} & Vehicle stunt simulator\newline \textbf{collaborate} & 1 \\
2 & {\normalsize\faGem} & Open-world heist game\newline \textbf{collect} & \cellcolor{goalAcolor}{\normalsize\faRunning} & \cellcolor{goalAcolor}\textbf{Wall-jump course}\newline \textbf{traverse} & \cellcolor{goalAcolor}{\normalsize\faRunning} & \cellcolor{goalAcolor}\textbf{Wall-jump drills}\newline \textbf{traverse} & \cellcolor{goalAcolor}4 &  & Knife duel game\newline -- & 5 \\
3 & {\normalsize\faCoins} & Open-world action RPG\newline \textbf{manage} &  & Chance-based game\newline -- & \cellcolor{goalAcolor}{\normalsize\faRunning} & \cellcolor{goalAcolor}\textbf{Staged wall-jump course}\newline \textbf{traverse} & \cellcolor{goalAcolor}>10 &  & Chance-based game\newline -- & 3 \\
4 & {\normalsize\faGem} & Fishing simulator\newline \textbf{collect} & \cellcolor{goalAcolor}{\normalsize\faRunning} & \cellcolor{goalAcolor}\textbf{Wall-jump drills}\newline \textbf{traverse} & \cellcolor{goalAcolor}{\normalsize\faRunning} & \cellcolor{goalAcolor}\textbf{Wall-jump tower}\newline \textbf{traverse} & \cellcolor{goalAcolor}>10 & {\normalsize\faHandsHelping} & Co-op flight survival\newline \textbf{collaborate} & 7 \\
5 & {\normalsize\faTrophy} & Team volleyball game\newline \textbf{compete} &  & Knife duel game\newline -- & \cellcolor{goalAcolor}{\normalsize\faRunning} & \cellcolor{goalAcolor}\textbf{Short wall-jump tower}\newline \textbf{traverse} & \cellcolor{goalAcolor}>10 & {\normalsize\faShield*} & Checkpoint roleplay\newline \textbf{defend} & >10 \\
\bottomrule
\end{tabular}
\caption{Case study of amplifying and suppressing the same concept ($\alpha=4$; feature 927 for \goalhl{goalAcolor}{\textit{traverse}}). `Before' denotes the rank of each game in the original top-10, and shading marks games labeled with the concept. Icons follow the concepts. Game titles and thumbnails are anonymized.}
\label{tab:amp_supp}
\end{table*}

\begin{table*}[t]
\centering
\scriptsize\hyphenpenalty=10000\exhyphenpenalty=10000
\setlength{\tabcolsep}{2pt}
\renewcommand{\arraystretch}{1.1}
\begin{tabular}{@{}c>{\centering\arraybackslash}m{0.5cm}>{\raggedright\arraybackslash}m{2.65cm}>{\centering\arraybackslash}m{0.5cm}>{\raggedright\arraybackslash}m{2.65cm}>{\centering\arraybackslash}m{0.5cm}>{\raggedright\arraybackslash}m{2.65cm}>{\centering\arraybackslash}m{0.7cm}>{\centering\arraybackslash}m{0.5cm}>{\raggedright\arraybackslash}m{2.65cm}>{\centering\arraybackslash}m{0.7cm}@{}}
\toprule
 & \multicolumn{2}{c}{Recent plays (oldest $\rightarrow$ newest)} & \multicolumn{2}{c}{Original top-5} & \multicolumn{3}{>{\columncolor{goalAcolor}[2pt][0pt]}c}{Suppressing \textit{compete for dominance}} & \multicolumn{3}{>{\columncolor{goalBcolor}[0pt][0pt]}c@{}}{Suppressing \textit{collect rare items}} \\
\cmidrule(lr){2-3}\cmidrule(lr){4-5}\cmidrule(lr){6-8}\cmidrule(l){9-11}
 & & Game, concept & & Game, concept & & Game, concept & Before & & Game, concept & Before \\
\midrule
1 & {\normalsize\faChessKnight} & Ability arena fighter\newline \textbf{tactical} & \cellcolor{goalBcolor}{\normalsize\faGem} & \cellcolor{goalBcolor}\textbf{Item-code hub}\newline \textbf{collect} & \cellcolor{goalBcolor}{\normalsize\faGem} & \cellcolor{goalBcolor}\textbf{Item-code hub}\newline \textbf{collect} & \cellcolor{goalBcolor}1 &  & Block-breaking game\newline -- & 6 \\
2 & \cellcolor{goalBcolor}{\normalsize\faGem} & \cellcolor{goalBcolor}\textbf{Item-code hub}\newline \textbf{collect} & {\normalsize\faChessKnight} & Reflex PvP game\newline \textbf{tactical} & {\normalsize\faChessKnight} & Reflex PvP game\newline \textbf{tactical} & 2 & \cellcolor{goalAcolor}{\normalsize\faTrophy} & \cellcolor{goalAcolor}\textbf{Platform brawler}\newline \textbf{compete} & \cellcolor{goalAcolor}5 \\
3 & \cellcolor{goalAcolor}{\normalsize\faTrophy} & \cellcolor{goalAcolor}\textbf{Platform brawler}\newline \textbf{compete} & \cellcolor{goalBcolor}{\normalsize\faGem} & \cellcolor{goalBcolor}\textbf{Collector hangout}\newline \textbf{collect} & \cellcolor{goalBcolor}{\normalsize\faGem} & \cellcolor{goalBcolor}\textbf{Collector hangout}\newline \textbf{collect} & \cellcolor{goalBcolor}3 & {\normalsize\faPuzzlePiece} & Physics puzzle\newline \textbf{solve} & >10 \\
4 &  & Block-breaking game\newline -- & \cellcolor{goalAcolor}{\normalsize\faTrophy} & \cellcolor{goalAcolor}\textbf{Platform fighting game}\newline \textbf{compete} & {\normalsize\faChessKnight} & Anime fighting sandbox\newline \textbf{tactical} & 8 & \cellcolor{goalAcolor}{\normalsize\faTrophy} & \cellcolor{goalAcolor}\textbf{PvP tower climb}\newline \textbf{compete} & \cellcolor{goalAcolor}>10 \\
5 & {\normalsize\faTheaterMasks} & Social utility game\newline \textbf{roleplay} & \cellcolor{goalAcolor}{\normalsize\faTrophy} & \cellcolor{goalAcolor}\textbf{Platform brawler}\newline \textbf{compete} & {\normalsize\faPuzzlePiece} & Story escape room\newline \textbf{solve} & 10 & {\normalsize\faRunning} & Precision jump course\newline \textbf{traverse} & >10 \\
\bottomrule
\end{tabular}
\caption{Case study of suppressing two concepts ($\alpha=4$; features 838 for \goalhl{goalAcolor}{\textit{compete}} and 2855 for \goalhl{goalBcolor}{\textit{collect}}). `Before' denotes the rank of each game in the original top-10, and shading marks games labeled with the suppressed concept. Icons follow the concepts. Game titles and thumbnails are anonymized.}
\label{tab:supp_contrast}
\end{table*}

\newpage
\begin{figure*}[t]
\begin{tcolorbox}[enhanced,colback=gray!4,colframe=black!65,colbacktitle=black!65,coltitle=white,
  fonttitle=\sffamily\bfseries\scriptsize,title={Prompt for Play Concept Generation},
  boxrule=0.5pt,arc=1.5pt,left=4pt,right=4pt,top=3pt,bottom=3pt,
  fontupper=\ttfamily\scriptsize\raggedright]
\#\# Context\\
The platform hosts user-generated games, such as simulators, obstacle courses, roleplay spaces, survival and horror games, shooters, RPGs, and strategy games. Each game is described by short text covering its gameplay, social experience, progression, and theme. The catalog already has these genre labels: \textcolor{blue!70!black}{\{genres\}}.

\medskip
\#\# Task\\
Propose about \textcolor{blue!70!black}{\{N\}} play purposes that cut across genres, covering social, progression, and creative dimensions as well as game mechanics. The definition of each concept will be turned into a text embedding and compared by cosine similarity with the embedding of each game's description. Games whose descriptions are semantically closest to a definition score highest for that concept, so each definition should read like the activity part of a game description.

\medskip
\#\# Requirements\\
- Express each concept as an activity or goal a player can pursue, not as a genre.\\
- Describe what players do positively, without negation or exclusions such as "not", "without", or "rather than".\\
- Make it judgeable from a game description alone and applicable to at least three genres.\\
- Do not include user-behavior conditions or infer personality, emotion, or psychological state.\\
- One game may match several concepts.\\
- Merge only near-duplicate concepts.

\medskip
\#\# Output Format\\
A JSON list of objects with five keys.\\
- label: a short label in plain words.\\
- definition: one or two sentences, the only text that will be embedded.\\
- matching\_examples and confusable\_non\_matching: matching and similar-looking non-matching activities, for human review only.\\
- example\_genres: exactly three genres from the list, each with a short activity example.\\
Do not invent real game names.
\end{tcolorbox}
\vspace{-2mm}
\caption{Prompt for generating play concepts. The name of platform and details are omitted for anonymity.}\label{fig:concept_prompt}

\end{figure*}

\begin{figure*}[t]
\begin{tcolorbox}[enhanced,colback=gray!4,colframe=black!65,colbacktitle=black!65,coltitle=white,
  fonttitle=\sffamily\bfseries\scriptsize,title={Prompt for Concept Judgment},
  boxrule=0.5pt,arc=1.5pt,left=4pt,right=4pt,top=3pt,bottom=3pt,
  fontupper=\ttfamily\scriptsize\raggedright]
\#\# Task\\
You review games on the platform. You are given twelve play concepts with definitions and the metadata of one game. Select every concept whose definition clearly describes what players do in this game. A game can match several concepts, one concept, or none.

\medskip
\#\# Input\\
Concepts: \textcolor{blue!70!black}{\{label: definition\}} for all twelve concepts.\\
Game: \textcolor{blue!70!black}{\{title, overview, core gameplay, social experience, progression and economy, distinctive hooks, world theme\}}.

\medskip
\#\# Rules\\
- matched: list every concept that applies, each with at least one evidence quote.\\
- none: the metadata is informative enough, and no concept applies.\\
- insufficient: the metadata is too thin or too vague to judge.\\
- For none or insufficient, traits must be empty.\\
- Copy concept labels exactly. Do not list a concept twice.\\
- Copy each quote character for character from the metadata.\\
- Judge only play characteristics observable in the text. Do not infer who the player is, what the player feels, or what the player intends.

\medskip
\#\# Output Format\\
One JSON object with status (matched, none, or insufficient) and traits, each with a concept label, evidence quotes, and a one-sentence explanation.
\end{tcolorbox}
\vspace{-2mm}
\caption{Prompt for judging concepts with LLMs. The judges see no concept scores, assigned concepts, or feature information. The name of platform and details are omitted for anonymity.}\label{fig:judge_prompt}
\vspace{-4mm}
\end{figure*}

\end{document}